\documentclass{ws-mpla}
\usepackage{psfig}
\begin{document}

\markboth{Saneesh Sebastian and V C Kuriakose}
{Scalar and Electromagnetic Quasinormal modes of Extended black hole in F(R) gravity}

\catchline{}{}{}{}{}

\title{Scalar and Electromagnetic Quasinormal modes of Extended black hole in F(R) gravity}

\author{\footnotesize  Saneesh Sebastian\footnote{
Typeset names in 8 pt Times Roman, uppercase. Use the footnote to 
indicate the present or permanent address of the author.}}

\address{Department of Physics, Cochin University of Science and
Technology, Kochi 682022, India\\
saneeshphys@cusat.ac.in}

\author{V C Kuriakose}

\address{Department of Physics, Cochin University of Science and
Technology, Kochi 682022, India\\
vck@cusat.ac.in
}

\maketitle

\pub{Received (Day Month Year)}{Revised (Day Month Year)}

\begin{abstract}
In this paper we study the scalar and electromagnetic perturbations of an extended black hole in F(R)
gravity. The quasinormal modes in two cases are evaluated and studied their behavior by plotting  
graphs in each case. To study the quasinormal mode, we use the third order WKB method. The present study 
shows that the absolute value of imaginary part of complex quasinormal modes increases in both cases, 
thus the black hole is stable against these perturbations. As the mass of the scalar field increases the imaginary
part of the frequency decreases. Thus damping slows down with increasing mass of the scalar field.

\end{abstract}
\ccode{PACS Nos.: 04.70.Dy, 04.70.-s }
\section{Introduction}	
	  One of the most exciting discoveries in cosmology after Hubble's discovery of expansion of universe  
is the accelerated expansion of the universe\cite{ar,pm}. In order to explain the acceleration, it becomes necessary
to modify the existing theory of gravitation proposed by Einstein, ie., the General Theory of Relativity. It can be done in
two ways, the first one is to modifying the energy-momentum part of the field equation. This necessitates the existence of an exotic component 
called dark energy. The cosmological constant is the simplest dark energy model. Since the effect of dark energy 
can be felt only at large scales there are no experimental supports to it till now. The second approach to explain the observed 
acceleration of the universe is to modify the curvature dependent part of field equation. This results in 
extended theories of gravity. There are a number of extended theories of gravity in literature. F(R) gravity, 
Gauss-Bonnet gravity, teleparallel gravity and new massive gravity are some examples of extended theories of gravity.
Among them, F(R) gravity is simple and successful\cite{beg}.

	Even though the F(R) gravity is most simple and comparatively successful, we cannot find out a unified F(R) model which explains 
both galaxy rotation curves of different galaxies as well as the accelerated expansion of the universe. Some F(R) models like $R^{-1}$
suffers the problem of ghosts. We must choose an F(R) model that must be stable for different types of perturbations. The different studies towards the 
quantum gravity show that there must be terms of curvature invariants but there are no curvature invariants in general 
F(R) theory\cite{tps}.

The action of F(R) gravity is obtained 
by replacing the Ricci scalar $R$ in the Einstein-Hilbert action by a general function of $R$. Varying the action thus obtained
with respect to the metric we get the F(R) field equation, it is a fourth order equation. Comparing the field equation 
of General Theory of Relativity which is second order, the field equation of F(R) gravity is more difficult to solve. Static cylindrically  symmetric 
interior solutions in F(R) gravity is studied by Sharif et al\cite{ms} and Godel solution is obtained by Santos\cite{afs} in F(R,T) gravity. Thermodynamics of 
evolving Lorentzian wormholes is studied  by Saiedi\cite{hs}.
In this work we use a solution originally obtained by Sebastiani et al\cite{ls}.
 Scattering of scalar field in this metric has been studied by the authors\cite{sv}.

      The quasinormal modes are the characteristic sound of black hole, which was first pointed out by Vishveshwara\cite{cv}.
Chandrasekhar and Detweller\cite{cd} studied the quasinormal modes of Schwarzschild black hole. Since then there are a number of 
works appeared in this field. For a review on quasinormal modes please see the references\cite{kdk,noll} and references therein. The study of quasinormal modes
becomes relevant because it 
 gives a direct way to detect black hole using the gravitational waves. There are a number of methods for evaluating quasinormal 
frequencies but the WKB method developed by Schutz, Will and Iyer is the simplest\cite{sw,iw,si}.

     The paper is organized as follows. In section 2 we discuss the static spherically symmetric solution in F(R) gravity In Section 3, we study
     the quasinormal modes of the extended black hole perturbed by electromagnetic field.  
In Section 4 we study the quasinormal modes of extended black hole using scalar perturbation for massive and massless cases.
WKB approximation is discussed in Section 5 and finally in Section 6 Conclusion of the present work is given.

\section{Static spherically symmetric solution in F(R) theory}
	  Starting with the F(R) action as 
\begin{equation}
S=\frac{1}{2\kappa^{2}}\int d^{4}x \sqrt{-g} F(R),
\end{equation}
where $g$ is the determinant of the metric and $F$ is the general function of Ricci scalar $R$. We start with a static spherically symmetric
solution.\cite{ls}
We take  $F(R)= \sqrt{R+6C_{2}}$ and this particular form of F(R) is taken as it gives a valid black hole solution. We take this F(R) as positive,
where $C_{2}$ is an integration constant. This is chosen either as positive or zero and all terms in the radical are taken as positive.
We write the static spherically symmetric solution as\cite{ls}
\begin{equation}
ds^{2}=-e^{2\beta(r)} B(r)dt^{2}+\frac{dr^{2}}{B(r)}+r^{2}d\Omega^{2}.
\end{equation}
With constant $\beta$, we get the solution as
\begin{equation}
ds^{2}=-B(r)dt^{2}+\frac{dr^{2}}{B(r)}+r^{2}d\Omega^{2},
\end{equation}
where $B(r)$ is given by
\begin{equation}
B(r)=1-\frac{C_{1}}{r^{2}}+C_{2}r^{2}.
\end{equation}
Comparing with Schwarzschild-de Sitter solution, the $C_{2}$ term in the solution which is the coefficient of $r^{2}$
can represent the cosmological constant term. For in this case we choose $C_{1}=2\alpha m$ and $C_{2}=0$. We choose
$C_{2}=0$ because we restrict
our study to a space time which is asymptotically flat\cite{jb}. We select $C_{1}$ as $2\alpha m$ such that the metric should be positively 
related to the mass only in such cases we get the correct Newtonian limit and $\alpha$ is chosen as a length parameter.
Thus the extended metric in F(R) gravity is given by\cite{ls,sv},   
\begin{equation}
ds^{2}=-\left(1-\frac{2\alpha m}{r^{2}}\right)dt^{2}+\left(1-\frac{2\alpha m}{r^{2}}\right)^{-1}dr^{2}+r^{2}d\theta^{2}+r^{2}sin^{2}\theta d\phi^{2},
\end{equation}
where $m$ is the black hole mass and $\alpha$ is a length parameter of the metric. The black hole mass is related to the metric linearly and thus
we assume a linear mass term with one length parameter. This length parameter can be adjusted to obtain an effective rotation curves of galaxies and 
gravitational lensing.  
F(R) is chosen as above because only this form of F(R) will give a static spherically symmetric black
hole solution. F(R) models with $R^{-1}$, $R^2$ etc.\cite{ft,star} have been studied but they do not possess black hole solution. 

\section{Evolution of quasinormal modes-Electromagnetic perturbation}
	We are studying the evolution of Maxwell field in this extended space time in F(R) gravity. The Maxwell's equation can be written as, 
\begin{equation}
 F^{\mu\nu}_{;\nu}=0,\ F_{\mu\nu}=A_{\nu,\mu}-A_{\mu,\nu},
\end{equation}
 where $F_{\mu\nu}$ is the electromagnetic field tensor and $A_{\mu}$ is the vector potential and vector potential $A_{\mu}$ can be expressed as 
 four dimensional vector spherical harmonics\cite{rr},
\begin{equation}
A_{\mu}(t,r,\theta,\phi)=\displaystyle\sum_{l,m}\left(\left[
\begin{array}{c}
0\\0\\ \frac{a^{lm}}{sin\theta}\partial_{\phi}Y_{lm}\\-a^{lm}(t,r)sin\theta \partial_{\phi} Y_{lm}
\end{array}\right]+\left[
\begin{array}{c}
f^{lm}(t,r) Y_{lm}\\h^{lm}(t,r) Y_{lm}\\k^{lm}(t,r)\partial_{\theta} Y_{lm}\\k^{lm}(t,r)\partial_{\phi} Y_{lm}
\end{array}\right]
\right),
\end{equation}
where $l$ and $m$ are angular momentum quantum number and azimuthal quantum number respectively. The first column has a parity of $(-1)^{l+1}$ and 
the second column has $(-1)^{l}$. We define the tortoise coordinates as
$\frac{dr_{*}}{dr}=\left(1-\frac{2\alpha m}{r^{2}}\right)^{-1}$ such that after some mathematical steps, we get the Regge-Wheeler\cite{rw}
equation as,
\begin{equation}
\frac{d^{2}}{dr_{*}^{2}}\Phi(r)+(\omega^{2}-V)\Phi(r)=0,
\end{equation}
with the potential $V$ given by, 
\begin{equation}
V(r)=\left(1-\frac{2\alpha m}{r^{2}}\right)\left(\frac{l(l+1)}{r^{2}}\right),
\end{equation}
\begin{table}[t]
\tbl{Quasinormal frequencies - electromagnetic perturbation.}
{\begin{tabular}{@{}cccccc@{}} \toprule
$l$ & $n$ & Re($\omega$) & Im($\omega$)\\
  \\ \colrule
2\hphantom{0}& 0 & 1.142700&\hphantom{0}-0.343905&\hphantom{0} \\
\hphantom{0}&\hphantom{0}1&\hphantom{0}0.992173&\hphantom{0}-1.082550& \hphantom{0} \\
\hphantom{0} & 2 & 0.767162\hphantom{0} & -1.877770\hphantom{0}  \\
\hphantom{0} & 3 & 0.475127\hphantom{0} & -2.699680\hphantom{0} \\
3\hphantom{0} & 0 & 1.67599\hphantom{0} & -0.348490\hphantom{0} \\
\hphantom{0} & 1 & 1.567820\hphantom{0} & -1.070310\hphantom{0} \\
\hphantom{0} & 2 & 1.389310\hphantom{0} & -1.835120\hphantom{0} \\
\hphantom{0} & 3 & 1.160460\hphantom{0} & -2.628530\hphantom{0} \\
\hphantom{0} & 4 & 0.881184\hphantom{0} & -3.440780\hphantom{0} \\
4\hphantom{0} & 0 & 2.19324\hphantom{0} & -0.350460\hphantom{0} \\
\hphantom{0} & 1 & 2.109240\hphantom{0} & -1.066020\hphantom{0} \\
\hphantom{0} & 2 & 1.961730\hphantom{0} & -1.812700\hphantom{0} \\
\hphantom{0} & 3 & 1.769050\hphantom{0} & -2.586900\hphantom{0} \\
\hphantom{0} & 4 & 1.536860\hphantom{0} & -3.380010\hphantom{0} \\
5\hphantom{0} & 0 & 2.70389\hphantom{0} & -0.351472\hphantom{0} \\
\hphantom{0} & 1 & 2.635290\hphantom{0} & -1.064070\hphantom{0} \\
\hphantom{0} & 2 & 2.510170\hphantom{0} & -1.799570\hphantom{0} \\
\hphantom{0} & 3 & 2.342660\hphantom{0} & -2.559400\hphantom{0} \\
\hphantom{0} & 4 & 2.140600\hphantom{0} & -3.338360\hphantom{0}  \\ \botrule
\end{tabular}\label{ta1}} 
\end{table}
\begin{figure}[h]
\centerline{\psfig{file=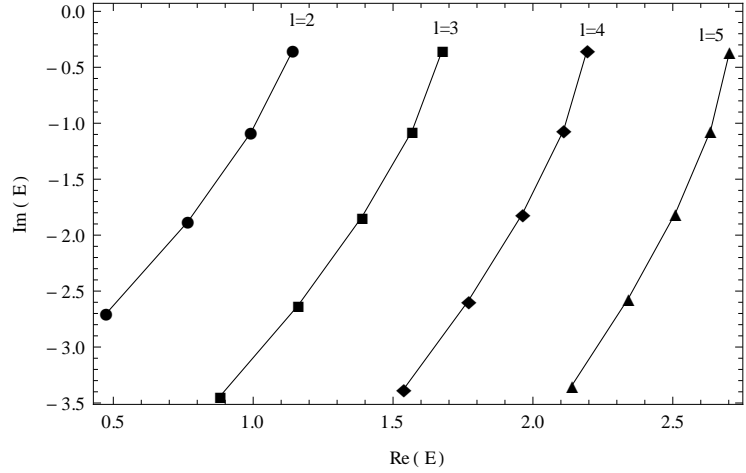,width=10cm}}
\vspace*{8pt}
\caption{Quasinormal modes - electromagnetic perturbation - massless case. \label{f1}}
\end{figure}
where $\Phi(r)=a^{lm}$ for parity $(-1)^{(l+1)}$ and $\Phi(r)=\frac{r^{2}}{l(l+1)}(-i\omega h^{lm}-\frac{df^{lm}}{dr})$  
for parity $(-1)^{l}$. Using WKB approximation method given in section.5 we can evaluate quasinormal modes of electromagnetic perturbation.
The electromagnetic quasinormal modes are given in Table. 1
\section{Quasinormal modes - Scalar perturbation}
In this section, we study the quasinormal modes of the extended black hole in F(R) gravity perturbed by scalar fields. The metric is given by Eq.(5). Since we are considering 
the scalar perturbation, we can use the Klein Gordon equation. We consider here both the massless
and massive scalar field. The Klein-Gordon equation for a massive scalar field is,
\begin{equation}
\Box\Psi-u^{2}\Psi=\frac{1}{\sqrt{-g}}\left(g^{\mu\nu}\sqrt{-g}\Psi_{,\mu}\right)_{,\nu}-u^{2}\Psi=0,
\end{equation}
where $u$ is the mass of the scalar field. In order to separate the wave function into radial, temporal, and angular 
parts, we write $\Psi$ as 
\begin{equation}
\Psi=\Phi(r) Y_{lm}(\theta,\phi) e^{-i\omega t}.
\end{equation}
Substituting Eq. 11 in Eq. 10 and after straight forward calculation, the radial part is given by
\begin{equation}
\left[\left(1-\frac{2\alpha m}{r^{2}}\right)^{-1}\omega^{2}+\left(1-\frac{2\alpha m}{r^{2}}\right)\partial_{r}^{2}-\frac{l(l+1)}{r^{2}}+\frac{1}{r}\partial_{r}\left(1-\frac{2\alpha m}{r^{2}}\right)-u^{2}\right]\Phi=0.
\end{equation}

Introducing the tortoise coordinate as $dr_{*}=\frac{dr}{1-\frac{2\alpha m}{r^{2}}}$, Eq.(8) can be written as,  
\begin{equation}
\left(\frac{d^{2}}{dr_{*}^{2}}+\omega^{2}-V(r)\right)\Phi(r)=0,
\end{equation}
where  the potential V(r) is given by 
\begin{equation}
V(r)=\left(1-\frac{2\alpha m}{r^{2}}\right)\left(\frac{l(l+1)}{r^{2}}+\frac{4\alpha m}{r^{4}}+u^{2}\right).
\end{equation}
$r_{*}$ ranges from $-\infty$ to$+\infty$
\begin{table}[t]
\tbl{Quasinormal frequencies - scalar perturbation.}
{\begin{tabular}{@{}cccccc@{}} \toprule
$l$ & $n$ & Re($\omega$) & Im($\omega$) & Re($\omega$) & Im($\omega$)\\
 & & $u=0$&$u=0$& $u=0.1$& $u=0.1$ \\ \colrule
2\hphantom{0}& 0 & 1.24391&\hphantom{0}-0.358055&\hphantom{0}1.24565 &\hphantom{0} -0.357272 \\
\hphantom{0}&\hphantom{0}1&\hphantom{0}1.09979&\hphantom{0}-1.12409& \hphantom{0}1.10026&\hphantom{0} -1.12267 \\
\hphantom{0} & 2 & 0.887537\hphantom{0} & -1.946290\hphantom{0} & 0.887124\hphantom{0}& -1.94536 \\
\hphantom{0} & 3 & 0.615554\hphantom{0} & -2.792610\hphantom{0} & 0.614869\hphantom{0}& -2.79230 \\
3\hphantom{0} & 0 & 1.74795\hphantom{0} & -0.355789\hphantom{0} & 1.749280\hphantom{0}& -0.355375 \\
\hphantom{0} & 1 & 1.64228\hphantom{0} & -1.09191\hphantom{0} & 1.64305\hphantom{0}& -1.09095 \\
\hphantom{0} & 2 & 1.46848\hphantom{0} & -1.87071\hphantom{0} & 1.46859\hphantom{0}& -1.86973 \\
\hphantom{0} & 3 & 1.24684\hphantom{0} & -2.67761\hphantom{0} & 1.24654\hphantom{0}& -2.67691 \\
\hphantom{0} & 4 & 0.977252\hphantom{0} & -3.50226\hphantom{0} & 0.976773\hphantom{0}& -3.50188 \\
4\hphantom{0} & 0 & 2.24908\hphantom{0} & -0.354872\hphantom{0} & 2.25015\hphantom{0}& -0.354617 \\
\hphantom{0} & 1 & 2.166320\hphantom{0} & -1.07915\hphantom{0} & 2.16710\hphantom{0}& -1.07849 \\
\hphantom{0} & 2 & 2.021180\hphantom{0} & -1.83435\hphantom{0} & 2.02154\hphantom{0}& -1.83355 \\
\hphantom{0} & 3 & 1.832020\hphantom{0} & -2.61689\hphantom{0} & 1.83203\hphantom{0}& -2.61616 \\
\hphantom{0} & 4 & 1.604590\hphantom{0} & -3.41799\hphantom{0} & 1.60436\hphantom{0}& -3.41743 \\
5\hphantom{0} & 0 & 2.74951\hphantom{0} & -0.35442\hphantom{0} & 2.75039\hphantom{0}& -0.354248 \\
\hphantom{0} & 1 & 2.68161\hphantom{0} & -1.072870\hphantom{0} & 2.68233\hphantom{0}& -1.07240 \\
\hphantom{0} & 2 & 2.55785\hphantom{0} & -1.814110\hphantom{0} & 2.55831\hphantom{0}& -1.81348 \\
\hphantom{0} & 3 & 2.39234\hphantom{0} & -2.579580\hphantom{0} & 2.39252\hphantom{0}& -2.57893 \\
\hphantom{0} & 4 & 2.19296\hphantom{0} & -3.364040\hphantom{0} & 2.19292\hphantom{0}& -3.36347 \\ \botrule
\end{tabular}\label{ta1}} 
\end{table}
\begin{figure}[pb]
\centerline{\psfig{file=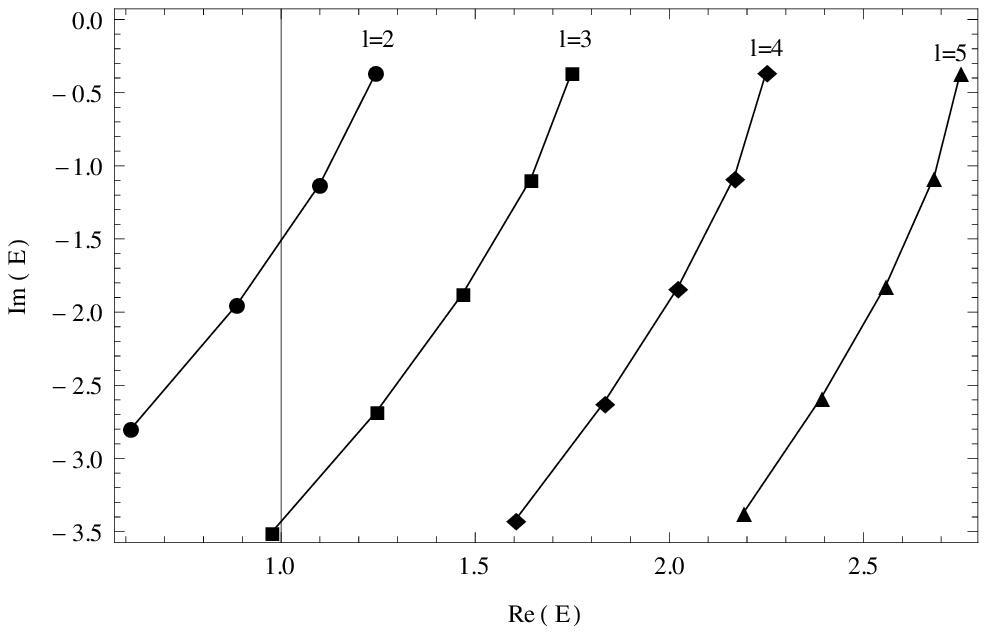,width=10cm}}
\vspace*{8pt}
\caption{Quasinormal modes - scalar perturbation - massive case. \label{f3}}
\end{figure}
\begin{figure}[pb]
\centerline{\psfig{file=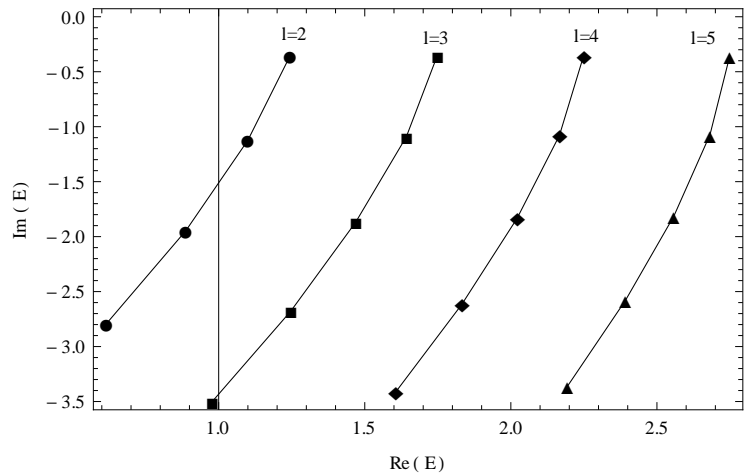,width=10cm}}
\vspace*{8pt}
\caption{Quasinormal modes - scalar perturbation - massless case. \label{f4}}
\end{figure}
The scalar field perturbation is studied and the corresponding quasinormal modes are obtained and are tabulated in Table. (2).
\section{WKB approximation method}
Using the third order WKB approximation method developed by Schutz and Will\cite{sw} and finally modified by Iyer and Will\cite{iw,si},
we can obtain the frequency of the quasinormal mode in general as, 
\begin{equation}
\omega^{2}=[V_{0}+(-2V''_{0})^{\frac{1}{2}}\Lambda]-i(n+\frac{1}{2})(-2V_{0}'')^{\frac{1}{2}}(1+\Omega),
\end{equation}
where
\begin{equation}
\Lambda=\frac{1}{(-2V_{0}'')^{\frac{1}{2}}}\left[\frac{1}{8}\left(\frac{V_{0}^{(4)}}{V_{0}''}\right)\left(\frac{1}{4}+\alpha^{2}\right)-\frac{1}{288}\left(\frac{V_{0}'''}{V_{0}''}\right)^{2}(7+60\alpha^{2})\right],
\end{equation}
and
\begin{eqnarray}
\Omega=\frac{1}{(-2V_{0}'')} [ \frac{5}{6912}\left(\frac{V_{0}'''}{V_{0}''}\right)^{4}\left(77+188\alpha^{2}\right)
-\frac{1}{384}\left(\frac{(V_{0}''')^{2}(V_{0}^{(4)})}{(V_{0}'')^{3}}\right)\left(51+100\alpha^{2}\right) \\ \nonumber
+\frac{1}{2304}\left(\frac{V_{0}^{(4)}}{V_{0}''}\right)^{2}(67+68\alpha^{2})+
\frac{1}{288}\left(\frac{V_{0}'''V_{0}^{(5)}}{(V_{0}'')^{2}}\right)(19+28\alpha^{2})\\ \nonumber
\frac{1}{288}\left(\frac{V_{0}^{(6)}}{V_{0}''}\right)(5+4\alpha^{2})],
\end{eqnarray}
where $\alpha=n+\frac{1}{2}$, and 
\begin{equation}
V_{0}^{(n)}=\frac{d^{n}V}{dr_{0}^{n}}|_{r_{*}=r_{*}max}.
\end{equation}
	
	The potentials given in Eq. 9 and Eq. 14 corresponding to electromagnetic and scalar perturbations are substituted in the WKB formula and obtained the 
corresponding complex frequencies. The quasinormal frequencies are tabulated in tables. In Table.1 we give the electromagnetic perturbed
quasinormal modes and in Table. 2 the quasinormal modes when the extended black hole is perturbed by the scalar fields. 
	
	We have plotted the data in Fig.(1), for electromagnetic field case
and from this figure we can see that the real part of quasinormal frequencies decreases with increasing mode number $n$ for a 
given angular momentum quantum number $l$. The Fig.(2) and Fig.(3) show scalar field perturbations for massless and massive field respectively.
In Fig.(1) Fig.(2) and Fig(3), the top dots are for $n=0$. The mode value of imaginary 
part of frequencies increases rapidly showing that the oscillation is damping and black hole is stable against electromagnetic and scalar
perturbations. The absolute value of imaginary frequency increases with mode number showing damping.

\section{Conclusion}
In this paper we have studied the quasinormal modes of extended black hole in F(R) gravity. The black hole space-time is perturbed by 
electromagnetic and scalar waves and behavior of the resulting quasinormal modes are evaluated. The present study shows that 
the imaginary part of complex quasinormal modes for both cases increase showing damping of oscillations. Thus the black hole is stable against
both scalar and electromagnetic perturbations. The damping time increases with increasing mass of the field.

\section*{Acknowledgments}
SS wishes to thank CSIR, New Delhi for financial support under CSIR-SRF
scheme. VCK is thankful to UGC, New Delhi for financial support through a 
Major Research Project and wishes to acknowledge Associateship of IUCAA, Pune,
India

\end{document}